\documentclass[runningheads]{mmef}\usepackage[]{latexsym}\usepackage{amssymb}\usepackage{amsmath}\usepackage{lscape}\usepackage{appendix}\usepackage{graphicx}\usepackage{amsfonts, wasysym}\input{psfig.sty}

\begin{document}\pagestyle{myheadings} \setcounter{page}{1}  \mainmatter\title{A Stochastic Model for the \\Analysis of Demographic Risk in\\Pay-As-You-Go Pension Funds}\titlerunning{A Stochastic Model for the Analysis of Demographic Risk $\ldots \quad$}

\author{Alessandro Fiori Maccioni\thanks{I wish to thank Prof. Ron W. Anderson and Prof. Alessandro Trudda for their precious comments on an earlier version of this paper. I am also greatly indebted to the CNPADC for providing the data and would finally like to thank Dr Francesco Arzilli, Dr Claudio Detotto, Dr Simonetta Falchi, Dr Roberta Melis, Ms Lisa O'Donoughue and the two anonimous referees for their useful suggestions. This work would have never been possible without the `Master and Back' program and the support to research provided by the Regione Autonoma della Sardegna.} \inst{}}

\institute{CRENoS - Centre for North-South Economic Research\\ and Dipartimento di Economia, Impresa e Regolamentazione,\\Universit\`a di Sassari,\\Via Torre Tonda n. 34 - 07100 Sassari, Italy\\\email{alex.fiori.maccioni@uniss.it}}

\maketitle\begin{abstract}
This research presents an analysis of the demographic risk related to future membership patterns in pension funds with restricted entrance, financed under a pay-as-you-go scheme. The paper, therefore, proposes a stochastic model for investigating the behaviour of the demographic variable `new entrants' and the influence it exerts on the financial dynamics of such funds. Further information on pension funds of Italian professional categories and an application to the \textit{Cassa Nazionale di Previdenza e Assistenza dei Dottori Commercialisti} are then provided.


\smallskip\noindent {\bf M.S.C. classification.} \textsc{\small 11K45, 60J10, 60J20, 62N02.}

\noindent {\bf J.E.L. classification.} \textsc{\small C15, C32, C53, G23, J11.}

\end{abstract}

\bigskip\section{Introduction}
The financial sustainability of a pension scheme depends not only on the time-length of the benefit to be paid, subject to longevity risk, but also on the correct quantification of the contributions to be received. The uncertainty related to future contributions primarily affects the retirement plans based on a pay-as-you-go financing system (PAYG), where pensions are directly funded by current employees' salary deductions. Thus, an ``intergenerational pact" compels young generations (composed by current and future contributors) to sustain the older age groups. For the financial self-sufficiency of such pension schemes, it is essential that in the long term there should be equilibrium between the number of pensioners and the number of workers. Due to the inversion of the production cycle, delivering financial equilibrium in the short-medium term via an increase in the number of new members can be harmful in the long run, when contributors in the preceding period will become pensioners. If the ratio contributors/pensioners was to decrease in the future, due to an increase of an older population, there would be \textit{ceteris paribus} an increase of financial burden for the pension system; thus creating a disequilibrium in the pension scheme and the risk of financial difficulties.

Public PAYG pension schemes are generally opened to different professional groups, in order to prevent financial unbalances due to a decline in a specific profession (and the consequent decline in its contributors). Therefore, variations in the number of contributors are mostly influenced by changes in the age structure of the population. On the other hand, private PAYG pension schemes can choose to admit only a homogeneous class of people (e.g. employees of a specific firm, workers with given professional qualification, etc.). For such ``closed" pension funds, the demographic risk related to the variable `new entrants' is relevant because changes in the job market may influence the number of contributors. This is the case of the self-administered PAYG pension funds of Italian professional categories.

Until the early nineties, social security and pension disbursement in Italy were publicly funded and administered. Nearly all citizens, regardless of their income, were entitled to a pension that allowed them to approximately maintain the same socio-economic status they enjoyed while employed. The right to a pension was generously guaranteed as prescribed by the article 38 of the Italian Constitution. 
In 1994 the Legislative Decree No. 509 was passed, calling for the privatisation of certain sectors of social security and pension administration. Any professional group organized as an Order/Board (such as lawyers, accountants, engineers, doctors, pharmacists, etc.) was to create and administer its own retirement fund. Members who work autonomously  would deposit portions of their incomes during their working years, and receive pensions upon retirement; meanwhile, members who work as employees would still be entitled to public pension cover. These new financial institutions, called \textit{Casse di Previdenza e Assistenza dei Liberi Professionisti}, have been no longer dependent on governmental assistance. The change meant that, should a given fund reach a negative balance, there would be no more financial backup  from the public finances; thus retirees would have no pensions available to them.

In this perspective, the present study addresses the evaluation of demographic risk related to the variable `new entrants' in PAYG pension systems. It starts with a brief review of recent literature on pension fund risk management (Sect.\,2) and continues with the mathematical formalization of the problem (Sect.\,3) and an application to the \textit{Cassa Nazionale di Previdenza e Assistenza dei Dottori Commercialisti} (Sects.\,4 and 5). Finally, conclusions are drawn (Sect.\,6).

\bigskip\section{Risk Management in Pension Funds: State of the Art}
Over the last few years a vast literature regarding management and regulation of risk in pension systems has developed. The main topics on quantitative research have been the development of stochastic models for longevity risk and global asset return. Recent financial scandals have also improved research on governmental regulations for life insurance institutions.

An introduction to longevity risk with a comprehensive literary review can be found in \cite{pit2} and \cite{pit3}. Rigorous analyses of mortality projections have been conducted by Lee and Carter in \cite{lee:car}, Benjamin and Pollard in \cite{ben:pol}, Benjamin and Soliman in \cite{ben:sol}, Haberman and Renshaw in \cite{hab:ren}, Lee in \cite{lee}, Olivieri in \cite{oli}, Thatcher et al. in \cite{tha:kan} and Olivieri and Pitacco in \cite{oli:pit5}. Joint analyses of both financial and longevity risks have been proposed by Olivieri and Pitacco in \cite{oli:pit2} and by Coppola et al. in \cite{cop:dil:sib}. The securitisation of mortality risk has been analysed by Lin and Cox in \cite{lin:cox} and by Cairns et al. in \cite{cai:bla:dow}. 

Several stochastic models for global asset return in pension funds have been proposed; see for example Parker in \cite{par}, Cairns and Parker in \cite{cai:par}, Blake et al. in \cite{bla:cai:dow} and \cite{bla:cai:dow2}. Mandl and Mazurova in \cite{man:maz} use spectral decomposition of stationary random sequences for assessing defined benefit pension schemes under randomly fluctuating rates of return and numbers of entrants. Haberman in \cite{hab} identifies a `contribution rate risk' and considers as stochastic components both rate of return and contribution rate. Gerrard et al. in \cite{ger:hab:vig} analyse the financial risk faced by members of defined contribution schemes both during the service period and after retirement.

Stochastic analyses of new entrants in private pension schemes have been proposed by Janssen and Manca in \cite{jan:man} and by Colombo and Haberman in \cite{col:hab}. Sinn in \cite{sin}, \cite{sin2} and Abio et al. in \cite{abi:mah:pat} consider the age structure of future national population as a prime risk factor in PAYG public pension systems. Angrisani et al. in \cite{ang:att:bia:var} propose a demographic model for studying the impact on PAYG pension systems of future developments of the population. Bianchi et al. in \cite{bia:rom:vag} conduce joint demographic and behavioural analyses via dynamic microsimulation to test the economic effects of pension reforms.

A vast literature on risk management policies has been developed following defaults on life insurance sector; see for example Plantin and Rochet in \cite{pla:roc}. A debated point is whether competition among pension funds and moral hazard can expose funds to excessive risks that are not compatible with their social not-speculative function; see for example McClurken in \cite{mcc}. Bader in \cite{bad} suggests that pension funds should avoid investing in specific sectors in the stock market. Ryan and Fabozzi in \cite{rya:fab} study the defaults of US pension funds due to actuarial losses and not to wrong portfolio investments. Trudda in \cite{tru2} shows that marginal increments in global asset return appear to strongly reduce the default probability of the pension fund of an Italian professional Order, thus generating an incentive to take superfluous risks in case of lacking of regulations. Otranto and Trudda in \cite{otr:tru} urge the need for a risk rating system for pension funds and propose a cluster analysis based on GARCH volatility of their rates of return.

\bigskip\section{A Model for the Evaluation of New Entrants to a Pension Fund with Restricted Entrance}
The Population-Education-Profession (\mbox{P-E-P}) model, that we propose here, is a discrete-time stochastic model for the estimation of the new entrants in a pension fund with restricted entrance, such as that of a professional category. The model is based on the study of variables related to the demographic evolution of the population, the development of university instruction and the attraction of the profession. It can be used, with appropriate simplifications, to forecast the entrants in any kind of pension scheme. To the best of our knowledge, it is the first stochastic model specifically designed for the estimation of new members of a professional category (and, subsequently, of its pension fund). An early version of the model and a deterministic application are proposed in \cite{fio} and \cite{tru2}.

The intergenerational patterns of employment in a given professional group depend on different specific variables, both demographic (trend in population, trend in study choices, etc.) and economic (appeal of the profession, appeal of the firm, expected income, etc.). Thus, for a correct estimation of the future contributors to such a ``closed" pension fund we should address the following questions: \begin{itemize}
\item What will the demographic evolution of the reference population be?
\item	What are the trends in the choices of study regarding the specific profession?
\item	What is the attraction of the profession (or of the firm, in case of corporate pension fund)?
\item	How is the admission to the pension fund regulated (e.g. elective/compulsory entrance)?
\end{itemize}

\begin{figure}[htbp]
\centering
  \includegraphics[width=11.8cm]{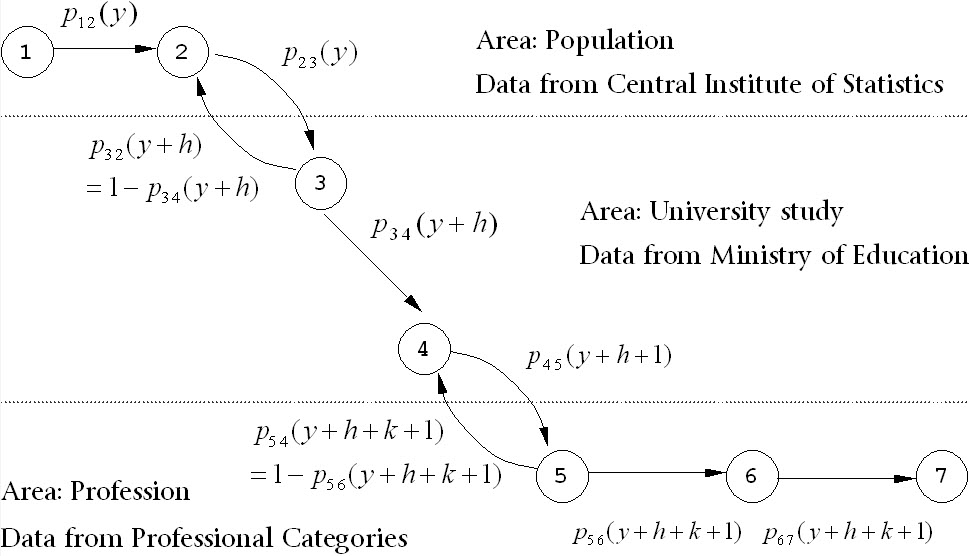}
	\caption{Markov chain for the estimation of future entrants to the pension fund of a professional category}
	\label{fig:example}
\end{figure}

The stages that a potential contributor has to leave behind before entering in the pension fund of a professional category have been represented in the Markov chain in Fig.\,\ref{fig:example}. \,Accordingly, we propose a model based on subsequent estimations of the population in different stages, as described in Table \ref{tab:stages}. \,The Markov chain is composed by the following states:  
\begin{enumerate}
\item Belonging to the cohort of reference (e.g. Italian population aged 18-25);
\item Having a high school diploma;
\item Being enrolled in the required course of study (e.g. Bachelor degree in Law);
\item Being graduated in the specific degree;
\item Starting the training period required for taking the admission exam;
\item Becoming a member of the professional category (e.g. becoming a Lawyer);
\item Joining the pension fund of the professional category.
\end{enumerate}
The $p_{ij}(t)$ in the Markov chain represents the probability of transition from state $i$ to state $j$ at time $t$; $h$ and $k$ are the expected lengths for successfully completing, respectively, the course of study and the professional training period. At time $t$, each potential future contributor can only be in one state. An individual can move to a greater state exclusively after fixed time periods ($0$, $h$ or $k$ time units) depending on the state itself.

\noindent \begin{table}[htbp]
\centering
\caption{Stages for estimating new entrants to the pension fund of a professional category}
\label{tab:stages}
\begin{tabular}{|l|l|}
\hline
\;Population of potential university students\: &  \:Regarding trends in population\: \\
\;$\qquad\qquad\qquad\qquad\downarrow$ & \\
\hline
\;Number of enrolments at university\: & \\
\;$\qquad\qquad\qquad\qquad\downarrow$ & \\
\;Enrolments in the specific field of study\: & \:Regarding trends in education\: \\
\;$\qquad\qquad\qquad\qquad\downarrow$ & \\
\;Graduations in the specific field of study\: & \\
\;$\qquad\qquad\qquad\qquad\downarrow$ & \\
\hline
\;Graduates who start the training period\: & \\
\;$\qquad\qquad\qquad\qquad\downarrow$ & \\
\;Trainees who complete the training period\: & \:Regarding the appeal of the profession\: \\
\;$\qquad\qquad\qquad\qquad\downarrow$ & \\
\;New members of the professional category\: & \\
\;\;\;= Trainees who pass the admission exam\: & \\
\;$\qquad\qquad\qquad\qquad\downarrow$ & \\
\hline
\;New entrants to the pension fund\: & \:Regarding entrance regulations\: \\
\hline
\end{tabular}
\end{table}

Let us introduce the stochastic process $POP(t)$ representing the population in state 1 at time $t$, which is, in other words, the starting population of the Markov chain, defined as: 
\begin{equation} \label{eqn:POP}POP(t)=\sum_{x=m}^{M} \overline{POP}_x(t) \; + \; \,\sigma_{pop}\,\epsilon \end{equation}
\noindent where  $\overline{POP}_x(t)$ is the expected value of the stochastic process $POP_x(t)$  representing the national population of age $x$ at time $t$;  \; $\sigma_{pop}$ is the standard deviation of $\sum{POP}_x(t)$; \;$m$ and $M$ are integer numbers indicating, respectively, the minimum and maximum age of the cohorts considered in the potential population; \; $\epsilon$  is a standard normal variable. Equation \ref{eqn:POP} holds for  $\sum{POP}_x(t)\:\geq \:-\sigma_{pop}\,\epsilon$ \; because population must be equal to a non-negative number.

Let $P_{ij}(t)$  be a stochastic process defined as:
\begin{equation} \label{eqn:Pij} P_{ij}(t)= \left\{ \begin{array}{lcl}
\,0 \quad & if \, & \epsilon <-\frac{\overline{p}_{ij}(t)}{\sigma_{ij}}\\
\overline{p}_{ij}(t) + \,  \sigma_{ij} \;\epsilon \quad & if \, & \epsilon \geq -\frac{\overline{p}_{ij}(t)}{\sigma_{ij}}
\end{array} \right.
\end{equation}
\noindent where $\overline{p}_{ij}(t)$  is the expected value of the probability of transition $p_{ij}(t)$, $\sigma_{ij}$  is its standard deviation, and $\epsilon$  is a standard normal variable. The $P_{ij}(t)$  can be seen as an approximation of the transition probability  $p_{ij}(t)$, but not as a probability in itself because it can assume values higher than one.

Let $NE(t)$  be a Markov process representing the number of new entrants to the pension fund at time $t$, defined as:
\begin{eqnarray} \label{eqn:NE} NE(t)=POP(t-h-k)P_{12}(t-h-k)P_{23}(t-h-k)\cdot \cr
\cr \cdot \,P_{34}(t-k)P_{45}(t-k)P_{56}(t)P_{67}(t) .\end{eqnarray}

To make the Markov chain more compliant with generally available statistics, we simplify it by merging some steps. Thus, we focus on the stochastic processes $P_{13}(t)$, $P_{34}(t)$, $P_{46}(t)$  and $P_{67}(t)$, and we rewrite (\ref{eqn:NE}) as:
\begin{equation} \label{eqn:NE2} NE(t)=POP(t-h-k)P_{13}(t-h-k)P_{34}(t-k)P_{46}(t)P_{67}(t) .\end{equation}

Equation (\ref{eqn:NE2}) represents the fundamental formula of the \mbox{P-E-P} Model. For sufficiently small values of the ratios  $\overline{p}_{ij}(t)/\sigma_{ij}$ and assuming the independence of the random processes, it also results:
\begin{eqnarray} NE(t)\approx[\sum_{x=m}^{M}\overline{POP}_x(t-h-k)+\sigma_{pop}\,\epsilon\,][\overline{p}_{13}(t-h-k)+\sigma_{13}\,\epsilon\,]\cdot \cr
\cr \cdot \,[\overline{p}_{34}(t-k)+\sigma_{34}\,\epsilon\,][\overline{p}_{46}(t)+\sigma_{46}\,\epsilon\,][\overline{p}_{67}(t)+\sigma_{67}\,\epsilon\,]\end{eqnarray}
\begin{eqnarray} E[NE(t)]\approx \sum_{x=m}^{M} \overline{POP}_x(t-h-k) \; \overline{p}_{13}(t-h-k) \; \overline{p}_{34}(t-k) \; \overline{p}_{46}(t) \; \overline{p}_{67}(t)\end{eqnarray}
and:
\begin{eqnarray} Var[NE(t)]\approx \{[\sum_{x=m}^{M} \overline{POP}_x(t-h-k)]^2+\sigma^2_{pop}\}[\overline{p}_{13}(t-h-k)^2+\sigma_{13}^2]\cdot \cr
\cr \cdot \,[\overline{p}_{34}(t-k)^2+\sigma_{34}^2][\overline{p}_{46}(t)^2+\sigma_{46}^2][\overline{p}_{67}(t)^2+\sigma_{67}^2] .\end{eqnarray}

In pension funds of professional categories, the number of future new entrants can be influenced by different variables, related to reference population, education choices, school completion, professional appeal and entrance regulations. The \mbox{Population-Education-Profession} model analyses such variables with five main stochastic processes.  Indeed, $POP(t)$  represents the national population who fulfil the university age requirements,   $P_{13}(t)$ the rate of diffusion of university studies among the population, $P_{34}(t)$   the success rate in university study, $P_{46}(t)$   the admission rate in the professional category, and $P_{67}(t)$   the inscription rate to the category's pension fund.

All of the parameters of the model can be estimated according to data that is easily available to professional categories, coming from the Central Institute of Statistic, from the Ministry of Education and from Professional Associations. Assuming that the period between $t-1$ and $t$ corresponds to the calendar year $y$, we can approximate $P_{13}(t)$  as the ratio between university enrolments (in the specific field of study) and the starting population at calendar year $y$. Accordingly, we can estimate $P_{34}(t)$  as the ratio between graduations at year $y$ and enrolments in previous year $y-h$;  $P_{46}(t)$ as the ratio between new members of the professional category at year $y$ and graduates at previous year $y-k$; $P_{67}(t)$  as the ratio between new entrants to the pension fund and new members of the category at year $y$. Such ratios can sometimes have values higher than 1 (e.g. because of university reforms, of changes in the fund's entrance regulations, etc.) and this is reflected in the definition of $P_{ij}(t)$.

The \mbox{P-E-P} model makes a breakthrough in the evaluation of new entrants to pension funds of professional categories. Classic actuarial techniques are mostly based on the time series analysis of past membership trends; this approach can underestimate the probability of sudden changes in the demographic dynamics, because it only considers the past outcome (new entrants time series) but not how this outcome had been generated. On the contrary, the \mbox{P-E-P} model can evaluate the risk of abrupt changes in the different variables that influence the number of future members (e.g. due to university reforms, decline in the professional appeal, etc.). Accordingly, it can assign different levels of demographic risk in funds that presented similar new entrants time series.

\bigskip\section{A Numerical Application of the \mbox{P-E-P} Model}

We used the \mbox{P-E-P} model for estimating the demographic evolution of the \textit{Cassa Nazionale di Previdenza e Assistenza dei Dottori Commercialisti} (CNPADC). The CNPADC is the private self-managed pension fund of Italian Chartered Accountants (\textit{Ordine dei Dottori Commercialisti}) set up by Law n. 335/1995. It is a cash-balance pension fund financed with a pay-as-you-go system, so the money collected during the year from contributors is immediately used for paying the same year's pensions, and only partially saved.

The demographic structure of the CNPADC is that of a ``young" retirement fund, meaning that it has still not reached the long-term natural balance between the number of contributors and pensioners. This is a consequence of the huge increase in new memberships that has occurred since the mid-Nineties. In the period 1995-2005, there has been an increase of more than 100\% in the total number of members, that have risen from 21,762 to 44,706; the ratio workers/pensioners has increased from 7 to 9.6; instead the ratio between contribution revenues and benefit expenses has been nearly constant, passing from 2.74 to 2.72.

The values of $\overline{POP}_x(t)$ for the years 1998-2006 (thus, influencing the number of new contributors in 2007-2015) have been taken from the official estimates of the Italian population published by the Italian Institute of Statistics (ISTAT).  The values of $\overline{POP}_x(t)$  and $\sigma_{pop}$  for the period 2007-2050 have been estimated on the basis of the National Demographic Forecasting of Italian population published by ISTAT.

The required degrees for being admitted to the Order of Chartered Accountants have been considered the \textit{Laurea} (\textit{triennale, specialistica} and \textit{vecchio ordinamento}) in Economics, Management and Business Administration. We focus on the period that has followed the main reform of the Italian university system in decades, which has introduced the three-years Bachelor degree.
The value of $h$, the average time for completing university has been considered equal to 5 years. The value of $k$, the average time for completing the professional training and passing the exam, has been considered equal to 4 years. We assume $m=18$ and $M=25$, basing such assumption on the fact that more than 95\% of university population in Italy is composed by students aged 18-25, and this ratio has been reasonably constant over time in the last two decades.

The moments of $P_{13}(t)$  and  $P_{34}(t)$ have been estimated according to historical data from ISTAT and the Italian Ministry of University (MIUR). The moments of  $P_{46}(t)$ and $P_{67}(t)$  have been estimated according to historical data from MIUR, CNPADC and \textit{Fondazione Aristeia}. For the evaluation of  $P_{67}(t)$  we have also considered historical data on cancellations from the fund, so that the estimates of new entrants can be considered net of cancellations. 

\begin{figure}[htbp]
\centering
  \includegraphics[height=4.5cm]{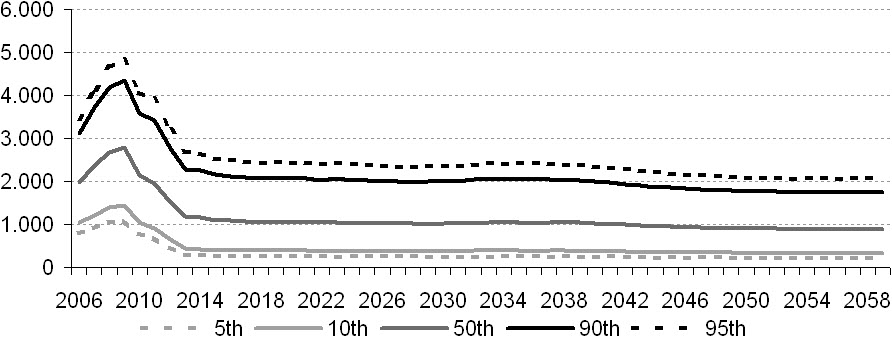}
	\caption{Total new entrants to the CNPADC pension fund, percentiles of the frequency distributions, years 2006-2059}
	\label{fig:total}
\end{figure}

\begin{figure}[htbp]
\bigskip  \includegraphics[height=4.5cm]{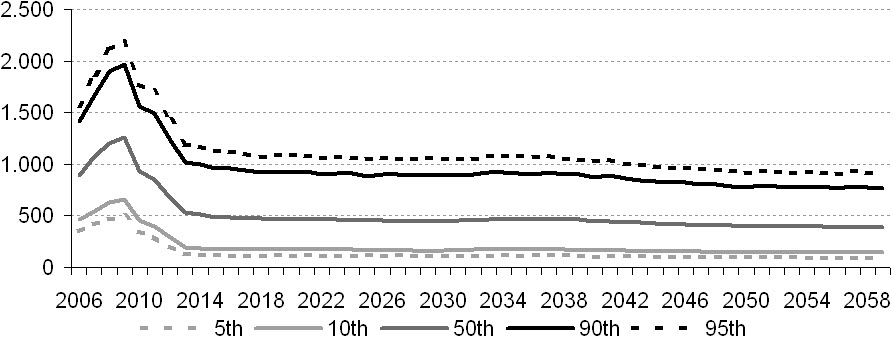}
	\caption{Female new entrants to the CNPADC pension fund, percentiles of the frequency distributions, years 2006-2059}
	\label{fig:females}
\end{figure}

\begin{figure}[htbp]
\bigskip
  \includegraphics[height=4.5cm]{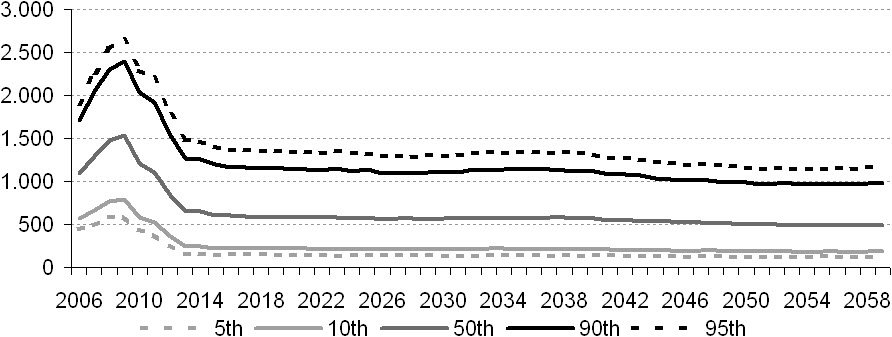}
	\caption{Male new entrants to the CNPADC pension fund, percentiles of the frequency distributions, years 2006-2059}
	\label{fig:males}
\end{figure}

Estimates of new entrants to the CNPADC fund have been made with the \mbox{P-E-P} model and tested with the Monte Carlo method. With 10,000 simulations, we have drawn the probability distribution of the future entrants, divided by sex, for the period 2006-2059. Tables \ref{tab:parameters} and \ref{tab:newentrants} present, respectively, the parameters of the model and the expected results. Figures \ref{fig:total}, \ref{fig:females} and \ref{fig:males} present the percentiles of the frequency distributions for new entrants obtained in the simulations (respectively: total, females and males).
\noindent \begin{table}[!htbp]
\centering
\caption{CNPADC pension fund: parameters of the P-E-P model}
\label{tab:parameters}
\begin{footnotesize}
\begin{tabular}{|l|cc|cc|cc|cc|}
\hline
\; & \; $\overline{p}_{1\,3}(t)$ \; & \; $\sigma_{1\,3}$ \; & \; $\overline{p}_{3\,4}(t)$ \; & \; $\sigma_{3\,4}$ \; & \; $\overline{p}_{4\,6}(t)$ \; & \; $\sigma_{4\,6}$ \; & \; $\overline{p}_{6\,7}(t)$ & $\sigma_{6\,7}$ \; \\
\hline 
\, Females \,&\, 0.0085 &\, 0.0007 \,&\, 0.5110 &\, 0.1996 \,&\, 0.0811 &\, 0.0291 \,&\, 0.6261 &\, 0.1088\, \\
\hline
\, Males \,&\, 0.0090 &\, 0.0005 \,&\, 0.5110 &\, 0.1996 \,&\, 0.0893 &\, 0.0320 \,&\, 0.6388 &\, 0.1108\, \\
\hline
\end{tabular}
\end{footnotesize}
\end{table}
\noindent \begin{table}[!htbp]
\centering
\caption{New members of the CNPADC fund, expected values, years 2006-2059}
\label{tab:newentrants}
\begin{scriptsize}
\begin{tabular}{|r|r r r|r|r|r r r|r|r|r r r|r|r|r r r|}
\cline{1-4}\cline{6-9}\cline{11-14}\cline{16-19}
Year	&	Mal.	&	Fem.	&	Tot.	&	\;	\qquad	&	Year	&	Mal.	&	Fem.	&	Tot.	&	\;	\qquad	&	Year	&	Mal.	&	Fem.	&	Tot.	&	\;	\qquad	&	Year	&	Mal.	&	Fem.	&	Tot.	\\
\cline{1-4}\cline{6-9}\cline{11-14}\cline{16-19}2006	&	1,119	&	915	&	2,034	&	\;	\qquad	&	2020	&	595	&	516	&	1,111	&	\;	\qquad	&	2034	&	592	&	511	&	1,103	&	\;	\qquad	&	2048	&	561	&	447	&	1,008	\\
2007	&	1,330	&	1,088	&	2,418	&	\;	\qquad	&	2021	&	592	&	513	&	1,105	&	\;	\qquad	&	2035	&	593	&	512	&	1,106	&	\;	\qquad	&	2049	&	556	&	443	&	999	\\
2008	&	1,509	&	1,235	&	2,744	&	\;	\qquad	&	2022	&	591	&	511	&	1,102	&	\;	\qquad	&	2036	&	593	&	512	&	1,105	&	\;	\qquad	&	2050	&	552	&	440	&	992	\\
2009	&	1,565	&	1,280	&	2,845	&	\;	\qquad	&	2023	&	589	&	510	&	1,099	&	\;	\qquad	&	2037	&	592	&	511	&	1,103	&	\;	\qquad	&	2051	&	549	&	437	&	986	\\
2010	&	1,280	&	976	&	2,256	&	\;	\qquad	&	2024	&	585	&	506	&	1,091	&	\;	\qquad	&	2038	&	590	&	508	&	1,098	&	\;	\qquad	&	2052	&	546	&	435	&	981	\\
2011	&	1,166	&	904	&	2,071	&	\;	\qquad	&	2025	&	582	&	503	&	1,085	&	\;	\qquad	&	2039	&	586	&	504	&	1,090	&	\;	\qquad	&	2053	&	544	&	434	&	978	\\
2012	&	896	&	724	&	1,621	&	\;	\qquad	&	2026	&	578	&	499	&	1,077	&	\;	\qquad	&	2040	&	578	&	497	&	1,076	&	\;	\qquad	&	2054	&	543	&	432	&	975	\\
2013	&	713	&	573	&	1,286	&	\;	\qquad	&	2027	&	576	&	497	&	1,072	&	\;	\qquad	&	2041	&	570	&	490	&	1,060	&	\;	\qquad	&	2055	&	542	&	432	&	974	\\
2014	&	650	&	561	&	1,210	&	\;	\qquad	&	2028	&	574	&	495	&	1,068	&	\;	\qquad	&	2042	&	562	&	484	&	1,046	&	\;	\qquad	&	2056	&	542	&	432	&	974	\\
2015	&	621	&	541	&	1,162	&	\;	\qquad	&	2029	&	574	&	495	&	1,068	&	\;	\qquad	&	2043	&	554	&	476	&	1,030	&	\;	\qquad	&	2057	&	542	&	432	&	974	\\
2016	&	613	&	533	&	1,146	&	\;	\qquad	&	2030	&	576	&	497	&	1,072	&	\;	\qquad	&	2044	&	546	&	470	&	1,016	&	\;	\qquad	&	2058	&	542	&	432	&	974	\\
2017	&	606	&	526	&	1,132	&	\;	\qquad	&	2031	&	579	&	500	&	1,078	&	\;	\qquad	&	2045	&	539	&	464	&	1,003	&	\;	\qquad	&	2059	&	543	&	433	&	976	\\
2018	&	600	&	521	&	1,121	&	\;	\qquad	&	2032	&	584	&	504	&	1,088	&	\;	\qquad	&	2046	&	574	&	458	&	1,032	&	\;	\qquad	&		&		&		&		\\
2019	&	597	&	518	&	1,114	&	\;	\qquad	&	2033	&	588	&	508	&	1,106	&	\;	\qquad	&	2047	&	568	&	452	&	1,020	&	\;	\qquad	&		&		&		&		\\
\cline{1-4}\cline{6-9}\cline{11-14}\cline{16-19}
\end{tabular}
\end{scriptsize}
\end{table}

Notably, the highest values of new members are in the period 2006-2012. This is an effect of the reform of the university system, started in the academic year 2000/01; students who did not previously complete their \textit{Laurea} (which required from 4 to 6 years, depending on the field of study) have been allowed to be re-enrolled to an equivalent post-reform Bachelor degree (\textit{Laurea triennale}) without having to re-sit for the exams that they had already passed. Thus, a large number of students, who had previously left university without completing their 4-6 years program, have quickly obtained a Bachelor degree. This phenomenon is reflected in the estimates of $P_{13}(t)$ and $P_{34}(t)$, which indicate that it will cease its effect from 2008 (thus, affecting the number of new contributors until 2012).

\bigskip\section{An Analysis of the Financial Dynamics of the CNPADC Fund}
\subsection{The Model}
The aim of this section is to estimate the effects of the variable `new entrants' on the financial dynamics of the CNPADC retirement fund in the period 2006-2046.

The fund value $V_t$, corresponding to the value of the net assets belonging to the fund at time $t$, has been modelled with the following recursive equation:
\begin{equation} V_t=V_{t-1} \cdot (1+r_t)+C_t+R_t-B_t-E_t \end{equation}
\noindent where $r_t$ is the nominal annual rate of return, $C_t$, $B_t$  and  $E_t$ represent respectively the amounts of contribution income, pension disbursement and administrative expenses generated in the period $[t-1,\:t]$. All of the cash flows are assumed to take place at the end of each period as this is more consistent with the fund's regulations.

\subsection{Contribution Income}
The annual contribution income at time $t$ has been estimated with the following equation:
\begin{eqnarray} C_t = \sum_{g=1}^{G}\,\,\, \sum_{s=1}^{S}\sum_{x= \alpha +1}^{\pi}\sum_{a=1}^{A} \,\,\, c_{g\,s\,x\,a}(t) \cdot N_{s\,x\,a}(t)\,,\qquad \qquad \qquad \cr
\cr \,
\qquad \qquad \qquad \forall (x,\,a) \in N\!\times \!N -\{ x>\hat{x}_{b\,s\,t} \, \wedge \, a\geq\hat{a}_{b\,s\,t} \}
\label{eqn:contrib} \end{eqnarray}
\noindent where $c_{g\,s\,x\,a}(t)$ is the average contribution of type $g$\, paid at time $t$ by an individual of sex $s$, age $x$ and seniority $a$,\, and $N_{s\,x\,a}(t)$ is the number of members of the fund alive at time $t$ of sex $s$, age $x$ and seniority $a$.\, The term $G$ represents the number of types of contributions, $S$ the number of sex categories, $\alpha$ and $\pi$ the minimum and maximum potential age of contributors, and $A$ the maximum potential seniority. The terms $\hat{x}_{b\,s\,t}$ and $\hat{a}_{b\,s\,t}$ represent the retirement requirements of age and seniority, in force at time $t$, for members of sex $s$ to be entitled  to a benefit of type $b$; thus, the $N_{s\,x\,a}(t)$ considered in equation (\ref{eqn:contrib}) are cohorts of active members. 

It also results:
\begin{equation} c_{g\,s\,x\,a}(t)= \gamma_{g\,x\,a\,t} \cdot R_{g\,s\,x\,a}(t)
\end{equation}
\noindent where $\gamma_{g\,x\,a\,t}$ and $R_{g\,s\,x\,a}(t)$ represent respectively the contribution rate and the expected income amount for the determination of the contribution of type $g$ due at time $t$ by individuals of sex $s$, age $x$ and seniority $a$.

In the application we consider two main types of contributions of the fund: the \textit{soggettivo} and the \textit{integrativo}, determined annually as shares of, respectively, professional income and sales subject to Value Added Tax.

\subsection{Pension disbursement}
The annual pension disbursement at time $t$ has been estimated with the following equation:
\begin{eqnarray} B_t=\sum_{d=1}^{D}\,\,\,\sum_{s=1}^{S}\sum_{x= \beta+1}^{\omega}\sum_{a=1}^{A}\,\,\,b_{d\,s\,x\,a}(t) \cdot N_{s\,x\,a}(t)\,,\qquad \qquad \qquad \cr
\cr \,
\qquad \qquad \qquad \forall (x,\,a) \in N\!\times \!N:\{ x>\hat{x}_{b\,s\,t} \, \wedge \, a\geq\hat{a}_{b\,s\,t} \}
\label{eqn:benefit} \end{eqnarray}
\noindent where $b_{d\,s\,x\,a}(t)$ is the average contribution of type $d$\, received at time $t$ by a pensioner of sex $s$, age $x$ and seniority $a$,\, and $N_{s\,x\,a}(t)$ is the number of members of the fund alive at time $t$ of sex $s$, age $x$ and seniority $a$.\, The term $D$ represents the number of types of benefits, $\beta$ and $\omega$ the minimum and maximum potential age of pensioners. The terms $\hat{x}_{b\,s\,t}$ and $\hat{a}_{b\,s\,t}$ represent the retirement requirements of age and seniority in force at time $t$; thus, the $N_{s\,x\,a}(t)$ considered in equation (\ref{eqn:benefit}) are cohorts of retired members.

In the application we consider the two main types of benefits of the fund: the \textit{pensione di vecchiaia} and the \textit{pensione unica contributiva}, computed with a pro-rata mechanism in accordance with CNPADC regulations as amended by the statutory reform of 2004.

\subsection{Mortality rate}
Let $q_{s\,x}(t)$ be the probability at time $t$ that an individual of sex $s$ and age $x$, still alive at time $t$, will die before time $t+1$, defined as:
\begin{equation}
q_{s\,x}(t)=\left\{ \begin{array}{lcl}
\,0 & \quad if & \quad \epsilon < -\frac{ \overline{q}_{s\,x}(t) }{\sigma_{s\,x}} \\
\,\overline{q}_{s\,x}(t)+ \sigma_{s\,x} \, \epsilon & \quad if & \quad -\frac{\overline{q}_{s\,x}(t) }{\sigma_{s\,x}} \leq \epsilon \leq \frac{1-\overline{q}_{s\,x}(t)}{\sigma_{s\,x}} \\
\,1 & \quad if & \quad \epsilon > \frac{1-\overline{q}_{{s\,x}}(t)}{\sigma_{s\,x}}
\end{array} \right.
\label{eqn:q}
\end{equation}
\noindent where $\overline{q}_{s\,x}(t)$ and $\sigma_{s\,x}$ represent respectively the expected value and the standard deviation of $q_{s\,x}(t)$, and $\epsilon$ is a standard normal variable.

We model $\overline{q}_{s\,x}(t)$ as:
\begin{equation} \label{eqn:expq} \overline{q}_{s\,x}(t)=(1+\mu_{s\,x})^{t-t_0} \cdot q_{s\,x}(t_0) \end{equation}
\noindent where $q_{s\,x}(t_0)$ and $\mu_{s\,x}$ represent respectively an initial known value and the expected annual rate of change of $q_{s\,x}(t)$.

The aim of the mortality model is to evaluate the impact of the accidental component of longevity risk on the financial dynamics of the CNPADC fund. Such component is due to random deviations from the expected mortality values. It is a simple stochastic model for which parameters can easily be estimated from official data published by national institutes of statistics. The discrete time approach has been preferred since the time unit in the application is the year.

\subsection{Rate of return}
Let $r_t$ be a stochastic process representing the annual nominal rate of return defined as:
\begin{equation} \label{eqn:return} r_t=\overline{r}_t+X_t
\end{equation}
\noindent where $\overline{r}_t$ is the expected nominal rate of return in the period $[t-1,\:t]$ and $X_t$ is an $AR(1)$ process defined as:
\begin{equation} X_t=\varphi X_{t-1}+ \sigma\,\epsilon  \end{equation}
\noindent with $-1 < \varphi < 1$, where $\varphi$ and $\sigma$ are the parameters of the process, and $\epsilon$ is a standard normal variable. The proposed model represents a discrete form of the Vasicek model, with properties of normality, stationarity, mean reversion, and finite variance. The prudential asset allocation of the CNPADC fund is compatible with such properties.

The approach based on the Vasicek model seems to be particularily suitable for describing the rate of return in first pillar pension funds such as the CNPADC. It takes into account the possibility of obtaining negative values, which is a desirable feature when modelling the rate of return. Indeed, pension funds of Italian professional Orders can occasionally suffer financial losses although their social not-speculative function. This has happened for example in 2008.

\subsection{Technical Assumptions}
The model has been employed with the following assumptions.

\bigskip
\noindent Demographic hypotheses:
\begin{itemize}
\item Effective population of pensioners and contributors on the 1st of January 2006, according to data from the CNPADC, divided by sex, age and seniority.
\item Future entrants determined according to the \mbox{P-E-P} model as stated in the previous section, age of entry 29.
\item Initial mortality rates $q_{s\,x}(t_0)$ equal to rates of the 2006 Italian mortality table published by ISTAT; values of $\mu_{s\,x}$  and $\sigma_{s\,x}$  estimated according to data from ISTAT on Italian mortality in the period 1981-2006.
\end{itemize}
Financial hypotheses:
\begin{itemize}
\item Fund's net assets equal to 2,067,793,989 euros on the 1st of January 2006, according to the 2005 financial report.
\item Administrative costs of year 2006 equal to 28,447,830 euros, appreciated in the following years at 5\% nominal annual rate according to technical assumptions in \cite{ang}.
\item Parameters for the estimation of the rate of return are  $X_0=0$, $\varphi=-0.612$  and  $\sigma=0.03667$, according to results obtained by Melis in \cite{mel} for Italian fixed income investment funds.
\item Inflation rate equal to Italian Government's expectations exposed in the budget \textit{DPEF 2007-2011}, thus equal to 2\% in 2006, to 1.7\% in 2007, to 2.1\% in 2008, to 1.9\% in 2009, and to 1.6\% in the following years.
\end{itemize}
Contributions:
\begin{itemize}
\item Two types of contributions (\textit{soggettivo} and \textit{integrativo}) determined in accordance with CNPADC regulations. New entrants exercise the statutory right of exemption from contributions for the first 3 years.
\item Annual professional incomes and VAT sales equal, for each cohort of same sex and age, to the effective average values registered in 2005, appreciated at inflation rate.
\item Subjective contribution rate equal to 10.7\%\footnote{Subjective contribution rate varies electively between 10\% and 17\% of annual professional income. In 2005 the average rate was 10.71\%.}  of annual professional income; sums paid under the defined-contribution scheme accrued at 3.4\% nominal annual rate.
\item Integrative contribution rate equal to 4\% in 2006-2010 and successively to 2\% of annual professional VAT sales, according to the statutory reform of 2004.
\end{itemize}

\newpage Pensions:
\begin{itemize}
\item Benefits paid to pensioners who retired before the 1st of January 2006 equal to the effective average values registered in 2006.
\item Two types of benefits (\textit{pensione di vecchiaia} and \textit{unica contributiva}) for pensioners who retire after the 1st of January 2006, determined with a pro-rata mechanism in accordance with CNPADC regulations as amended by the statutory reform of 2004.
\item All benefits appreciated annually at the inflation rate. Each cohort of contributors retires immediatly after fulfilling the requirements. We do not consider benefit reversion to survivors.
\end{itemize}

\bigskip

\subsection{Results}
The probabilistic structure of the fund value has been estimated with stochastic simulation based on Monte Carlo techniques. This approach generates a range of outcomes which represents a probability distribution, conditional on the assumptions made. The number of outcomes has been 10,000 for each test made.

\begin{figure}[htbp]
\centering
  \includegraphics[width=10.8cm]{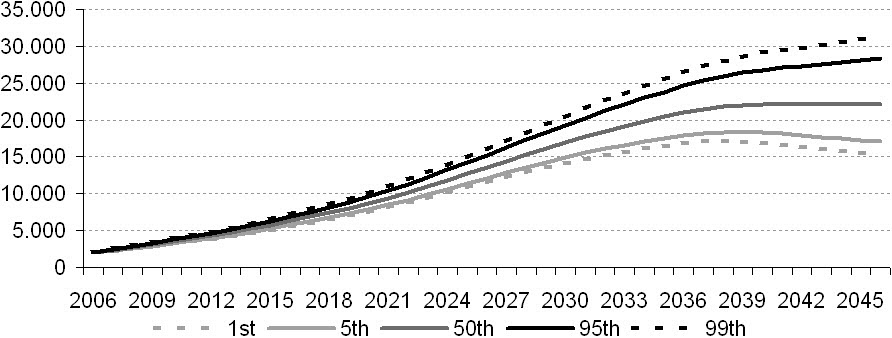}
	\caption{Value of the CNPADC pension fund in million euros, percentiles of the frequency distributions, years 2006-2046}
	\label{fig:fundvalue}
\end{figure}

In the first test, the three main risk factors --- mortality, new entrants and rate of return --- are considered as stochastic variables. Results are presented in Fig. \ref{fig:fundvalue} and indicate that, with 99.9\% confidence level, the CNPADC fund will maintain a positive value in the forecasting period.
The probability distribution of the fund value is nearly standard, slightly leptokurtic and right-skewed. The values of skewness and kurtosis tend to increase with the passing of time, as demonstrated in Tab. \ref{tab:freqtable}.

\begin{figure}[htb]
\centering
  \includegraphics[width=10.8cm]{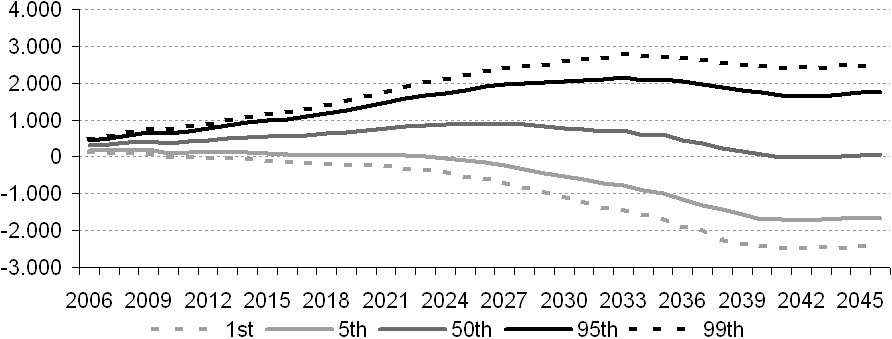}
	\caption{Total balance of the CNPADC pension fund in million euros, percentiles of the frequency distributions, years 2006-2046}
	\label{fig:totalbalance}
\end{figure}

\begin{figure}[htb]
\centering
  \includegraphics[width=10.8cm]{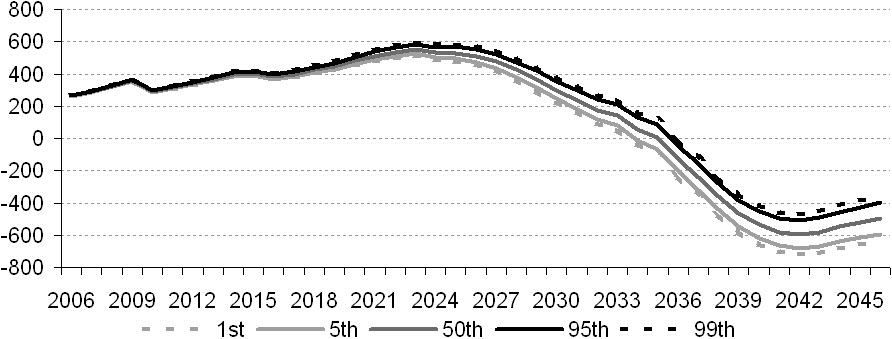}
	\caption{Pension balance of the CNPADC pension fund in million euros, percentiles of the frequency distributions, years 2006-2046}
	\label{fig:pensionbalance}
\end{figure}

The probability distribution of the total balance presents a large variation; its median value reaches a peak of 900 millions in 2026 and then decreases, reaching its minimum, 15 millions, in 2042, as demonstrated in Fig. \ref{fig:totalbalance}.
The percentile distribution of the pension balance indicates that this value will turn negative in the period between 2033 and 2037, with 99.9\% confidence level, because of the ageing of the population; its probability distribution presents a low dispersion around the median values, as demonstrated in Fig. \ref{fig:pensionbalance}. Returns on investments are expected to partially cover the increase in pension disbursement, thus preventing from abrupt slumps in the total balance.

We have conducted three other tests in which each risk factor --- mortality, rate of return and new entrants --- is considered as stochastic variable while the others are assumed deterministic.
Finally, one last simulation has estimated the expected financial dynamics of the fund considering all variables as deterministic. Results have been used to conduct a sensibility analysis of the fund, with the following conclusions.

The effects on the financial dynamics of the fund of random deviations from given mortality trends (that is, the accidental component of longevity risk) are negligible, and account for less than 1\% of dispersion around the median value. Indeed, this is a pooling risk and its effects tend to disappear in large pension funds such as the CNPADC.

\begin{figure}[htbp]
\centering
  \includegraphics[width=10.8cm]{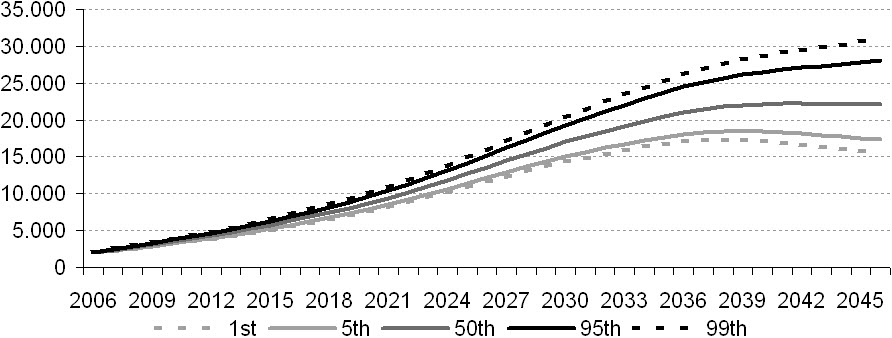}
	\caption{Value of the CNPADC pension fund in million euros, obtained considering stochastically only the variable `rate of return', percentiles of the frequency distributions, years 2006-2046}
	\label{fig:fundvalue3}
\end{figure}

Most of the variance in the fund value distribution seems to be described by the stochastic behaviour of the rate of return. This can be inferred by examining the similarities between the percentile distribution obtained in the first test, with all the three risk factors considered as stochastic, and the percentile distribution obtained under the hypotheses of stochastic rate of return and deterministic mortality and new entrants. The values are presented in Figs. \ref{fig:fundvalue} and \ref{fig:fundvalue3}.

\begin{figure}[htbp]
\centering
  \includegraphics[width=10.8cm]{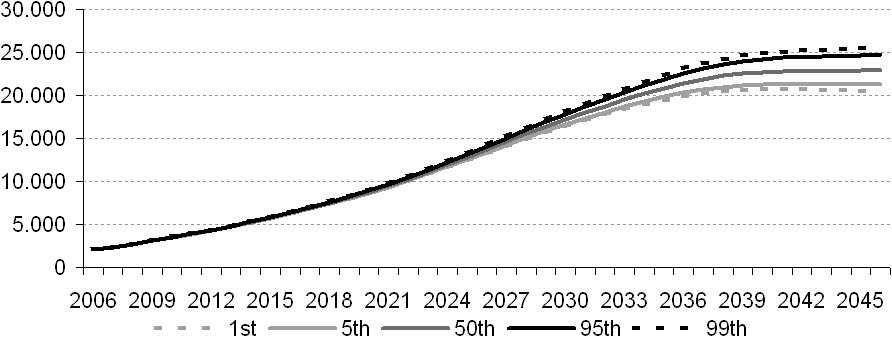}
	\caption{Value of the CNPADC pension fund in million euros, obtained considering stochastically only the variable `new entrants', percentiles of the frequency distributions, years 2006-2046}
	\label{fig:fundvalue2}
\end{figure}

Finally, the CNPADC fund seems to have a relatively low exposure to the risk related to future new entrants. This is suggested by the large impact of the rate of return on the percentile distribution of the fund value. Nonetheless, the stochastic behaviour of new entrants describes almost all of the variation in the pension fund balance. This can be deduced by examining the similarities between the variance in the pension balance distribution, obtained in the first test, and the fund value distribution obtained under the hypotheses of stochastic new entrants and deterministic mortality and rate of return. The values are presented in Figs. \ref{fig:fundvalue2} and \ref{fig:pensionbalance}.

\bigskip

\section{Conclusions}

In the present paper we have addressed the issue of the demographic risk related to future membership patterns in retirement funds with restricted entrance, financed under a pay-as-you-go scheme.

We have proposed a discrete-time Markov model for the estimation of new entrants in pension funds of professional categories, that highlights the interactions between demographic, economic and regulatory variables. The model considers the effects of trends in population, trends in education choices, appeal of the profession, and entrance regulations.

Numerical applications of the model have analysed the demographic and financial dynamics of the pension fund of Italian Chartered Accountants.
Demographic results have revealed the effects of a main reform in the university system on the number of new entrants to the pension fund. Financial results suggest that the fund has a relatively low exposure to the risk related to future new entrants, and that its main risk factor is the rate of return. Instead, the risk of random deviations from expected mortality trends generates negligible effects because of the large population of the fund.

Including other risk factors constitutes a main area of interest for further extensions. Specifically, the risk of regular deviations from expected mortality trends (that is, the systematic component of longevity risk) could be included.
\noindent \begin{table}[htbp]
\centering
\caption{Value of the CNPADC fund: moments of the frequency distributions obtained in the Monte Carlo simulation}
\label{tab:freqtable}
\begin{tabular}{|l|r|r|r|r|}
\hline
\;	Date:	&	1st Jan.\,2010	\;	&	1st Jan.\,2015	\;	&	1st Jan.\,2020	\;	&	1st Jan.\,2025	\;	\\
\;	Avg. Value:	&	\;3,525,456,641	\;	&	\;5,768,926,483	\;	&	\;8,725,501,938	\;	&	\;12,728,412,509	\;	\\
\;	St. Deviation:	&	143,505,878	\;	&	310,161,641	\;	&	542,057,540	\;	&	873,711,092	\;	\\
\;	Skewness:	&	0.096	\;	&	0.130	\;	&	0.196	\;	&	0.226	\;	\\
\;	Kurtosis:	&	0.054	\;	&	0.052	\;	&	0.119	\;	&	0.127	\;	\\
\hline
\hline
\;	Date:	&	1st Jan.\,2030	\;	&	1st Jan.\,2035	\;	&	1st Jan.\,2040	\;	&	1st Jan.\,2045	\;	\\
\;	Avg. Value:	&	\;17,045,796,388	\;	&	\;20,532,458,231	\;	&	\;22,259,140,016	\;	&	\;22,292,121,288	\;	\\
\;	St. Deviation:	&	1,337,223,372	\;	&	1,930,828,426	\;	&	2,597,537,506	\;	&	3,271,388,475	\;	\\
\;	Skewness:	&	0.237	\;	&	0.305	\;	&	0.358	\;	&	0.365	\;	\\
\;	Kurtosis:	&	0.116	\;	&	0.198	\;	&	0.266	\;	&	0.234	\;	\\
\hline
\end{tabular}
\end{table}

\noindent \begin{table}[htbp]
\centering
\caption{Expected cash flows of the CNPADC pension fund under deterministic assumptions, values in thousand euros, years 2006-2046}
\label{tab:cashflows}
\begin{scriptsize}
\begin{tabular}{|r| r| r| r| r| r| r| r| r| r|}
\hline
  &	\multicolumn{1}{c|}{Value at} &	\multicolumn{1}{c|}{Subjective} &	\multicolumn{1}{c|}{Integrative} &		\multicolumn{1}{c|}{Pension} &	\multicolumn{1}{c|}{Pension} &	\multicolumn{1}{c|}{Investm.} &	\multicolumn{1}{c|}{Admin.} &	\multicolumn{1}{c|}{Total} &	\multicolumn{1}{c|}{Value at} \\
 	\multicolumn{1}{|c|}{Year} &	\multicolumn{1}{c|}{1st January} &	\multicolumn{1}{c|}{Contrib.} &	\multicolumn{1}{c|}{Contrib.} &	\multicolumn{1}{c|}{Disburs.} &	\multicolumn{1}{c|}{Balance} &	\multicolumn{1}{c|}{Returns} &	\multicolumn{1}{c|}{Costs} &	\multicolumn{1}{c|}{Balance} &	\multicolumn{1}{c|}{31st Dec.} \\
  &	\multicolumn{1}{c|}{\textsc{A}}   &	\multicolumn{1}{c|}{\textsc{B}}   &	\multicolumn{1}{c|}{\textsc{C}}   &	\multicolumn{1}{c|}{\textsc{D}}   &	\multicolumn{1}{c|}{\textsc{E=\tiny{B+C-D}}}   &	\multicolumn{1}{c|}{\textsc{F}}   &	\multicolumn{1}{c|}{\textsc{G}}   &	\multicolumn{1}{c|}{\textsc{H=\tiny{E+F-G}}}    &	\multicolumn{1}{c|}{\textsc{I=\tiny{A+H}}}  \\
 \hline 
	2006	&	2,067,794	&	235,721	&	155,133	&	126,378	&	264,476	&	70,305	&	28,448	&	306,333	&	2,374,127	\\
	2007	&	2,374,127	&	251,469	&	165,874	&	127,616	&	289,727	&	80,720	&	29,870	&	340,577	&	2,714,704	\\
	2008	&	2,714,704	&	273,516	&	180,726	&	129,717	&	324,525	&	92,300	&	31,364	&	385,461	&	3,100,165	\\
	2009	&	3,100,165	&	296,535	&	195,999	&	132,946	&	359,588	&	105,406	&	32,932	&	432,062	&	3,532,227	\\
	2010	&	3,532,227	&	320,021	&	105,717	&	135,779	&	289,959	&	120,096	&	34,578	&	375,476	&	3,907,703	\\
	2011	&	3,907,703	&	344,246	&	113,646	&	143,858	&	314,034	&	132,862	&	36,307	&	410,588	&	4,318,291	\\
	2012	&	4,318,291	&	369,640	&	121,979	&	151,338	&	340,281	&	146,822	&	38,123	&	448,980	&	4,767,272	\\
	2013	&	4,767,272	&	395,262	&	130,491	&	158,308	&	367,446	&	162,087	&	40,029	&	489,504	&	5,256,776	\\
	2014	&	5,256,776	&	420,800	&	139,148	&	162,362	&	397,586	&	178,730	&	42,030	&	534,286	&	5,791,061	\\
	2015	&	5,791,061	&	442,274	&	146,518	&	183,278	&	405,514	&	196,896	&	44,132	&	558,278	&	6,349,340	\\
	2016	&	6,349,340	&	449,171	&	149,275	&	216,779	&	381,667	&	215,878	&	46,338	&	551,206	&	6,900,546	\\
	2017	&	6,900,546	&	465,270	&	155,035	&	223,675	&	396,630	&	234,619	&	48,655	&	582,593	&	7,483,139	\\
	2018	&	7,483,139	&	486,811	&	162,532	&	224,363	&	424,980	&	254,427	&	51,088	&	628,319	&	8,111,457	\\
	2019	&	8,111,457	&	499,487	&	167,154	&	220,442	&	446,199	&	275,790	&	53,643	&	668,346	&	8,779,803	\\
	2020	&	8,779,803	&	519,236	&	174,026	&	212,218	&	481,044	&	298,513	&	56,325	&	723,233	&	9,503,036	\\
	2021	&	9,503,036	&	535,622	&	179,714	&	204,108	&	511,228	&	323,103	&	59,141	&	775,190	&	10,278,226	\\
	2022	&	10,278,226	&	550,441	&	184,774	&	199,133	&	536,082	&	349,460	&	62,098	&	823,443	&	11,101,670	\\
	2023	&	11,101,670	&	562,260	&	188,751	&	197,256	&	553,755	&	377,457	&	65,203	&	866,009	&	11,967,678	\\
	2024	&	11,967,678	&	550,194	&	184,773	&	202,953	&	532,014	&	406,901	&	68,463	&	870,452	&	12,838,130	\\
	2025	&	12,838,130	&	555,862	&	186,599	&	212,552	&	529,909	&	436,496	&	71,886	&	894,519	&	13,732,649	\\
	2026	&	13,732,649	&	556,398	&	186,627	&	231,957	&	511,068	&	466,910	&	75,480	&	902,497	&	14,635,146	\\
	2027	&	14,635,146	&	553,372	&	185,563	&	260,129	&	478,807	&	497,595	&	79,255	&	897,147	&	15,532,293	\\
	2028	&	15,532,293	&	540,360	&	180,946	&	295,673	&	425,633	&	528,098	&	83,217	&	870,514	&	16,402,807	\\
	2029	&	16,402,807	&	528,597	&	177,013	&	338,932	&	366,678	&	557,695	&	87,378	&	836,995	&	17,239,802	\\
	2030	&	17,239,802	&	514,022	&	172,243	&	386,708	&	299,557	&	586,153	&	91,747	&	793,963	&	18,033,766	\\
	2031	&	18,033,766	&	500,528	&	167,916	&	429,731	&	238,712	&	613,148	&	96,334	&	755,526	&	18,789,292	\\
	2032	&	18,789,292	&	488,378	&	164,259	&	477,128	&	175,508	&	638,836	&	101,151	&	713,193	&	19,502,485	\\
	2033	&	19,502,485	&	507,559	&	170,759	&	533,357	&	144,961	&	663,084	&	106,209	&	701,837	&	20,204,322	\\
	2034	&	20,204,322	&	482,169	&	162,691	&	587,370	&	57,490	&	686,947	&	111,519	&	632,918	&	20,837,240	\\
	2035	&	20,837,240	&	485,158	&	163,733	&	642,333	&	6,557	&	708,466	&	117,095	&	597,928	&	21,435,168	\\
	2036	&	21,435,168	&	443,015	&	147,875	&	720,059	&	-129,169	&	728,796	&	122,950	&	476,677	&	21,911,845	\\
	2037	&	21,911,845	&	418,296	&	139,557	&	799,093	&	-241,240	&	745,003	&	129,097	&	374,666	&	22,286,511	\\
	2038	&	22,286,511	&	391,102	&	130,357	&	885,043	&	-363,585	&	757,741	&	135,552	&	258,605	&	22,545,115	\\
	2039	&	22,545,115	&	364,139	&	121,175	&	952,225	&	-466,911	&	766,534	&	142,330	&	157,293	&	22,702,409	\\
	2040	&	22,702,409	&	348,623	&	115,898	&	999,893	&	-535,372	&	771,882	&	149,446	&	87,064	&	22,789,472	\\
	2041	&	22,789,472	&	337,174	&	111,961	&	1,030,520	&	-581,385	&	774,842	&	156,919	&	36,539	&	22,826,011	\\
	2042	&	22,826,011	&	333,977	&	110,795	&	1,041,047	&	-596,274	&	776,084	&	164,764	&	15,046	&	22,841,057	\\
	2043	&	22,841,057	&	336,503	&	111,606	&	1,029,905	&	-581,796	&	776,596	&	173,003	&	21,797	&	22,862,854	\\
	2044	&	22,862,854	&	339,549	&	112,603	&	1,000,520	&	-548,368	&	777,337	&	181,653	&	47,316	&	22,910,170	\\
	2045	&	22,910,170	&	343,496	&	113,914	&	977,384	&	-519,974	&	778,946	&	190,735	&	68,236	&	22,978,407	\\
	2046	&	22,978,407	&	347,699	&	115,319	&	960,813	&	-497,795	&	781,266	&	200,272	&	83,199	&	23,061,606	\\
\hline
\end{tabular}
\end{scriptsize}
\end{table}

\end{document}